\documentclass[twoside,superscriptaddress,twocolumn,a4paper,floatfix,aps,longbibliography]{revtex4-2}

\usepackage{graphicx}
\usepackage[T1]{fontenc}
\usepackage{siunitx}
\usepackage{xcolor}

\usepackage{amsmath}

\usepackage{tabularx}

\usepackage{blkarray}
\usepackage{multirow}
\usepackage{braket}

\def\CP{{\rm CP}}
\def\NP{{\rm NP}}
\def\REEP{{\rm REEP}}

\def\<{\langle}
\def\>{\rangle}

\newcommand{\RHO}[1]{{\rho_{\rm #1}}}     
\newcommand{\SIGMA}[1]{{\sigma_{\rm #1}}} 

\usepackage{url}
\usepackage[colorlinks]{hyperref}
\hypersetup{%
    plainpages=true,
    breaklinks=true,
    hypertexnames=false,
    pageanchor=true,
    colorlinks=true,
    linkcolor={blue},
    citecolor={red},
    urlcolor={blue},
    anchorcolor={black}
    }

\begin{document}

 \title{Experimental relative entanglement potentials
 of single-photon states}

\author{Josef Kadlec} 
\affiliation{Palacký University in Olomouc, Faculty of Science,
Joint Laboratory of Optics of Palacký University and Institute of
Physics AS CR, 17. listopadu 12, 771 46 Olomouc, Czech Republic}

\author{Karol Bartkiewicz}
\affiliation{Institute of Spintronics and Quantum Information,
Faculty of Physics, Adam Mickiewicz University, 61-614 Poznań,
Poland}

\author{Antonín Černoch}
\affiliation{Institute of Physics of the Academy of Sciences of
the Czech Republic, Joint Laboratory of Optics of Palacký
University and Institute of Physics AS CR, 17. listopadu 50a, 772
07 Olomouc, Czech Republic}

\author{Karel Lemr}
\affiliation{Palacký University in Olomouc, Faculty of Science,
Joint Laboratory of Optics of Palacký University and Institute of
Physics AS CR, 17. listopadu 12, 771 46 Olomouc, Czech Republic}

\author{Adam Miranowicz}
\affiliation{Institute of Spintronics and Quantum Information,
Faculty of Physics, Adam Mickiewicz University, 61-614 Poznań,
Poland}

\begin{abstract}
Entanglement potentials (EPs) enable the characterization and
quantification of the nonclassicality of single-mode optical
fields by measuring the entanglement generated through beam
splitting. We experimentally generated single-photon states and
tomographically reconstructed the corresponding two-qubit states
to determine EPs defined via popular two-qubit measures of
entanglement. These include the potentials for the relative
entropy of entanglement (REEP), concurrence, and negativity. Among
our experimental states, we found those that are very close (at
least for some ranges of parameters) to the theoretical upper and
lower bounds on relative EPs (or relative nonclassicality), i.e.,
when one EP is maximized or minimized for a given value of another
EP. We experimentally confirmed the counterintuitive theoretical
result of Ref. [Phys. Rev. A \textbf{92}, 062314 (2015)] that the
relative nonclassicality (specifically, the negativity potential
for given values of the REEP) of single-photon states can be
increased by dissipation.
\end{abstract}

\date{\today}

\maketitle

\section{Introduction}

\begin{figure}[t]
\begin{center}
{\includegraphics[scale=0.5]{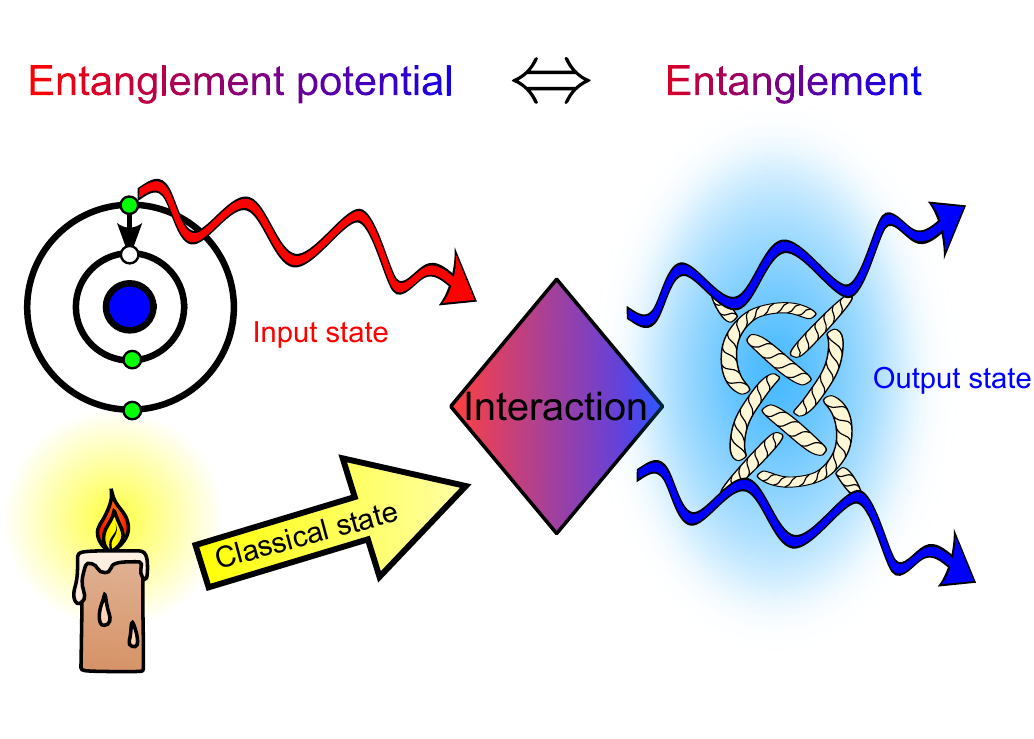}}\hspace*{0pt}\\ %
\caption{Conceptual scheme of entanglement potentials. A tested
state interacts with a purely classical state. The entanglement
detected in the output two-mode state reveals the nonclassicality
of the tested state and corresponds to its entanglement potential,
as quantified by a chosen entanglement measure.}\label{fig_scheme}
 \end{center}
 \end{figure}

Nonclassical states of light play a key role in various quantum
technologies, ranging from quantum communications and information
processing to quantum metrology. Their properties facilitate
inherently secure cryptography~\cite{Bennett_2014, Shor2000,
Ekert1991}, supremacy of quantum computing~\cite{Nielsen2010}, and
imaging and sensing with resolutions above classical limits
\cite{Fonseca1999, Jacobson1995}. An effective and experimentally
suitable approach for the nonclassicality description is, thus, an
indispensable ingredient enabling these technologies.

A quantum-optical state is considered nonclassical if its
Glauber-Sudarshan $P$-function is not positive semidefinite
\cite{glauberbook}. Based on this definition various measures of
single-mode nonclassicality have been proposed and studied, such
as: (i) the nonclassical distance~\cite{Hillery1987}, defined as
the distance of a given state to the closest classical state, (ii)
the nonclassical depth~\cite{Lee1991, Luetkenhaus1995}, defined as
the minimum amount of Gaussian noise required to convert a given
nonclassical state into a classical one, and (iii) entanglement
potentials (EPs)~\cite{Asboth2005}. The latter are studied in
detail in this paper. For a comparative study of these
nonclassicality measures see Ref.~\cite{Miranowicz2015b}.
Moreover, non-universal criteria (often referred to as
nonclassicality witnesses) are often applied in quantifying
single- and multimode nonclassicality. These include the
nonclassical volume~\cite{Kenfack2004}, the Wigner
distinguishability~\cite{Mari2011}, and the potentials for quantum
steering and Bell nonlocality~\cite{Miranowicz2023}, among dozens
of other criteria (see, e.g.,~\cite{Miranowicz2010} and references
therein). Here we apply the EPs, which have a number of advantages
compared to other methods: (i) some of the EPs can be easily
calculated (as shown in the following sections) in contrast to,
e.g., the nonclassical distance, (ii) they can distinguish the
nonclassicality degree of a wider variety of the classes of
states, compared to the nonclassical depth, (iii) they are
measurable, as we have recently demonstrated
experimentally~\cite{Kadlec2024}, and (iv) they are universal
\cite{Asboth2005}, as required for good nonclassicality measures,
in contrast to nonclassicality witnesses. Thus, EPs prove to be a
very convenient and experimentally feasible tool for describing
and quantifying nonclassicality. This concept is built upon the
ability of a nonclassical state to produce entanglement by
interacting with purely classical states, usually the vacuum (see
Fig.~1). An example of this effect is the single-photon state that
transforms into a Bell state after interacting with the vacuum on
a balanced beam splitter (BS).

A practical benefit of EPs is that they exploit the well developed
formalism for the quantification of bipartite entanglement to
describe single-mode nonclassicality. Moreover, any entanglement
measure might be used to define the corresponding EP, thus,
allowing us to gain a deeper understanding of nonclassicality, by
studying the mutual relations between various EPs. In this work,
we focus on the EPs defined by the Wootters
concurrence~\cite{Wootters1998}, the Peres-Horodecki
negativity~\cite{Vidal2002, Horodecki09review}, and the relative
entropy of entanglement (REE)~\cite{Vedral1997}.

This paper studies EPs, so it is related to our previous
theoretical~\cite{Miranowicz2015b, Miranowicz2015, Miranowicz2023}
and experimental~\cite{Kadlec2024} works. However, we should
emphasize that the results and their physical implications
reported here are different from those described in our related
experimental works~\cite{Kadlec2024}, which focused on the
relative potentials for entanglement, Einstein-Podolsky-Rosen
(EPR) steering, and Bell nonlocality. In contrast to that work, we
report here experimental relative EPs. Thus, this paper presents
an experimental demonstration validating the theoretical
predictions outlined in Refs.~\cite{Miranowicz2015,
Miranowicz2015b}, while our previous experiment reported
in~\cite{Kadlec2024} confirmed the concepts and predictions of
Ref.~\cite{Miranowicz2023}.

Specifically, we experimentally generate maximally and minimally
nonclassical single-photon states by quantifying their
nonclassicality with relative EPs, and refer to them as the
relative nonclassicality measures of a single photon. By
considering two EPs (say, EP1 and EP2) defined via different
measures of entanglement, we refer to the maximal (minimal)
relative EP1 vs. EP2, if EP1 is maximized (minimized) with respect
to a given value of EP2 for arbitrary single-qubit states. We
experimentally demonstrate that we can increase the relative EPs
for some entanglement measures by dissipation, thus experimentally
confirming the predictions of Ref.~\cite{Miranowicz2015}, and some
predictions of Refs.~\cite{Miranowicz2015b, Miranowicz2004,
Miranowicz2004b, Horst2013} on the bounds of relative entanglement
measures.

The measured value of a given EP is affected by both the genuine
properties of a tested state and the imperfections in its
interaction with the chosen classical state. In this paper, we
show that by analyzing mutual relations between several EPs
defined by standard entanglement measures, one can distinguish
between these two effects. This capability makes the EPs a
considerably more reliable tool for experimental nonclassicality
quantification.

We support our theoretical predictions~\cite{Miranowicz2015} with
a proof-of-principle experiment on the platform of linear optics.
In this experiment, we first generate the vacuum and one-photon
superposition (VOPS) states that then interact with the vacuum on
a BS. To avoid the need for the experimentally demanding vacuum
detection, while reconstructing the output state, we encode the
single-photon state into the horizontal polarization of a photon
and the vacuum into its vertical polarization, which can be
treated as a place-holder. We utilize the fact, that a vertically
polarized photon state has a vacuum component in its horizontal
mode, i.e., $\ket{0}_H\ket{1}_V$, where the numbers denote the
Fock states in the horizontal and vertical modes, respectively.
For a more detailed description of the experimental
implementation, see Sec.~III.

\section{Nonclassicality quantified by entanglement potentials}

\subsection{Entanglement potentials}

We calculate EPs for VOPS states, which can be described by the
density matrix
\begin{equation}\label{sigmapx}
\sigma(p,x) =
\begin{pmatrix}
1-p & x\\
x^* & p
\end{pmatrix}
\end{equation}
expressed in the basis of the vacuum, $\ket{0}$, and the Fock
single-photon state $\ket{1}$, with $p$ being the single-photon
probability and $x$ being a coherence term. Alternatively, one can
write $x = e^{i\phi}Dx_\text{max}$, where $x_\text{max}=
\sqrt{(1-p)p}$ is the maximal possible absolute value of the
coherence term, $D\in [0,1]$ is a dephasing factor, and $\phi \in
[0,2\pi)$ is an arbitrary phase. The dephasing factor can be
interpreted by assuming a pure state affected by a dephasing
channel, with a phase-flip probability of $f$, then $D = |1-2f|$.

Assuming a given VOPS state at one port of a BS and the vacuum
state at the other port, the output density matrix
reads~\cite{Miranowicz2023}
\begin{equation} \label{rhoqrpx}
\rho_{wr}(p,x) =
\begin{pmatrix}
1-p & -wrx & wtx &0\\
-wrx^* & pr^2 & -pw^2rt &0\\
wtx^* & -pw^2rt & pt^2& 0\\
0&0&0&0
\end{pmatrix},
\end{equation}
given in the standard computational basis
$\{\ket{00},\ket{01},\ket{10},\ket{11}\}$, where $r$ and $t$ are
the reflection and transmission coefficients, respectively.
Because the phases of $r$ and $t$ do not affect any reasonable
entanglement measure, we assume them both to be real non-negative.
We assume that both output ports of the BS are uniformly affected
by phase damping (see Appendix A.f), as described by the
phase-damping parameter $\kappa\equiv \kappa_1=\kappa_2$ or,
equivalently, the parameter $w=\sqrt{1-\kappa}$ for
$w,\kappa\in[0,1]$, where $w=1$ corresponds to a perfectly
coherent interaction.

The matrix in Eq.~(\ref{rhoqrpx}) can be used to calculate any
entanglement measure. In this paper, we use the negativity, the
concurrence, and the REE to quantify the output state
entanglement.

The negativity~\cite{Zyczkowski1998, Zyczkowski1999, Vidal2002},
which is related to the Peres-Horodecki entanglement criterion, is
defined for a two-qubit state $\rho$ as either zero or the
absolute value of the smallest negative eigenvalue of a partially
transposed density matrix and multiplied for convenience by a
factor of 2:
\begin{equation}
\text{N}(\rho) = \text{max}[0,-2\text{min eig}(\rho^\Gamma)],
\end{equation}
where the superscript $\Gamma$ denotes partial transposition.

The Wootters concurrence can be calculated using the
formula~\cite{Wootters1998}
\begin{equation}
\text{C}(\rho) =  \text{max}(0,2\lambda_{\text{max}} -
\sum_j\lambda_j),
\end{equation}
where $\lambda_j^2 = \text{eig}[\rho(Y\otimes Y)\rho^*(Y\otimes
Y)]$, $\lambda_\text{max} = \text{max}_j\lambda_j$, the asterisk
(*) denotes complex conjugation and $Y$ is a Pauli operator.

The REE~\cite{Vedral1997} is another popular two-qubit
entanglement measure and is defined as
\begin{equation}
 \text{REE}(\rho) = S(\rho||\rho^{\text{opt}}_\text{sep}) \equiv \underset{\rho_{\text{sep}} \in \mathcal{D}}{\text{min}} S(\rho||\rho_\text{sep}),
\end{equation}
in terms of the Kullback-Leibler distance,
$S(\rho||\rho_\text{sep}) = \text{Tr}(\rho\log_2\rho -
\rho\log_2\rho_\text{sep})$, of a given two-qubit state $\rho$
from the closest separable state $\rho^{\text{opt}}_\text{sep}$,
where $\mathcal{D}$ is the set of all two-qubit separable states
$\rho_\text{sep}$.

Thus, one can define the negativity potential (NP), the
concurrence potential (CP), and the relative entropy of
entanglement potential (REEP), as
\begin{align}
\text{NP}_{wr}[\sigma(p,x)]& = \text{N}[\rho_{wr}(p,x)],\\
\text{CP}_{wr}[\sigma(p,x)] &= \text{C}[\rho_{wr}(p,x)],\\
\text{REEP}_{wr}[\sigma(p,x)]& = \text{REE}[\rho_{wr}(p,x)],
\end{align}
respectively. In particular, these generalized EPs reduce to the
standard ones assuming balanced BS (with $t=r$) and no damping
($w=1$). Note that the potential is attributed to the single-mode
state $\sigma$ by evaluating the entanglement of the two-mode
state $\rho$.

To acquire a `true' EP of the input state one needs to have a
balanced ($r=t=1/\sqrt{2}$) and lossless ($w=1$) BS, otherwise the
values of the observed EPs are diminished due to imperfect
interactions. Simultaneously, such imperfect interactions can
increase relative EPs, thus they can be considered a resource, as
explained in detail below.

\subsection{Relative entanglement potentials}

We refer to the \emph{relative entanglement potentials} and
\emph{relative nonclassicality} of a single-mode state, when
considering an EP for a given value of another EP calculated for
the same state. Specifically, we study the following relative EPs:
the CP vs. NP, CP vs. REEP, and NP vs. REEP, and vice versa, as
shown in Fig. 2.

The yellow regions in all the panels of Fig.~2 show the allowed
values of the relative EPs for arbitrary single-photon states
assuming a perfectly balanced ($t=r$) and lossless BS. As
predicted theoretically in Ref.~\cite{Miranowicz2015}, the cyan
regions in Fig. 2(c) can be reached by relative EPs only for
imperfect BSs, e.g., assuming their phase damping, amplitude
damping, or unbalanced splitting.

Note that a straight line is observed for pure states when
comparing the CP and NP, as shown in Fig. 2(a). However, the REEP
is defined by logarithms, while the CP and NP are not. Thus,
straight lines are not observed for pure states in Figs. 2(b) and
2(c).

This work reports our experimental generation and tomographic
detection of: (i) the states in the cyan regions of Fig. 2(c) and
(ii) the states close to the maximally nonclassical states, which
are explicitly defined in Appendix~A.

\begin{figure*}
\begin{center}
\centering
\includegraphics[width=0.3\linewidth]{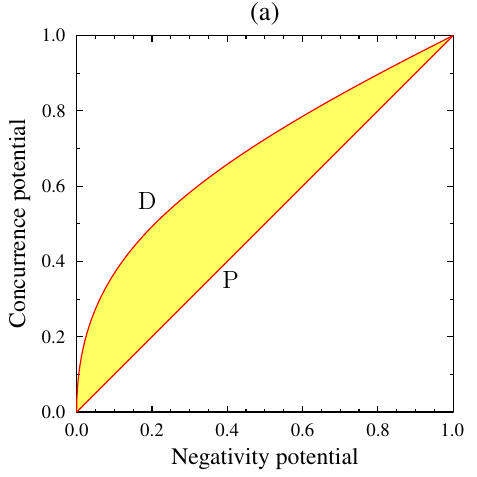}
\includegraphics[width=0.3\linewidth]{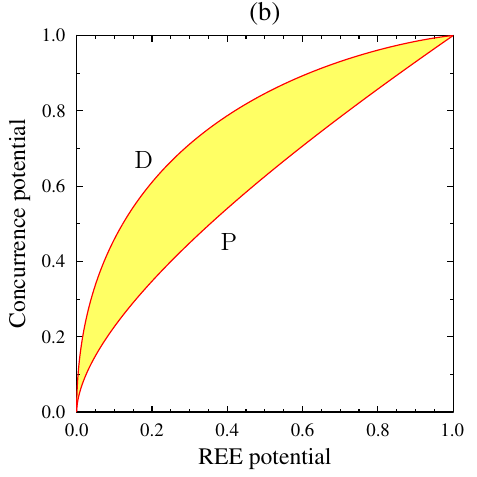}
\includegraphics[width=0.3\linewidth]{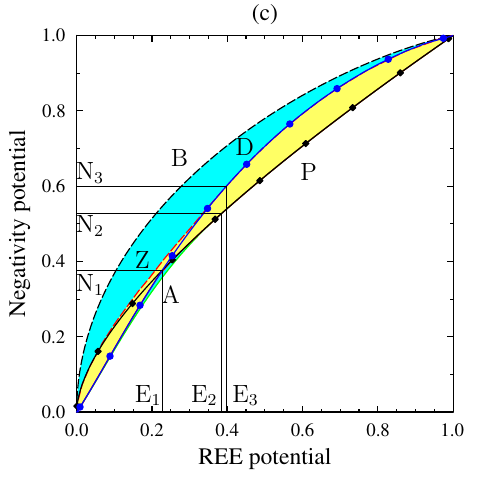}
\caption{Relative entanglement potentials:
 (a) $\CP_{wr}(\sigma)$ vs. $\NP_{wr}(\sigma)$,
 (b) $\CP_{wr}(\sigma)$ vs. $\REEP_{wr}(\sigma)$, and
 (c) $\NP_{wr}(\sigma)$ vs. $\REEP_{wr}(\sigma)$ for arbitrary single-qubit
states $\sigma$. Note that the two-qubit states $\rho_{wr}$, which
can be generated from $\sigma$ by a balanced BS ($t=r=1/\sqrt{2}$)
and assuming no damping ($w=1$) in a perfect EP-detection setup,
are located in the yellow regions. By contrast to this, the cyan
regions in (c) indicate those $\rho_{wr}$ which cannot be
generated from $\sigma$ in this way; their generation requires an
additional resource, i.e., a tunable BS with $t\neq r$, amplitude
or phase damping ($w<1$). Characteristic points are plotted at:
$(E_1\approx 0.228,N_1\approx 0.377)$, $(E_2 \approx 0.385,N_2
\approx 0.527)$, and $(E_3 \approx 0.397, N_3 \approx 0.6)$. Their
meaning is explained in the main text.}\label{fig1}
 \label{relative_EPs}
 \end{center}
\end{figure*}

\section{Experimental setup}
\begin{figure}
\begin{center}
{\includegraphics[scale=0.85]{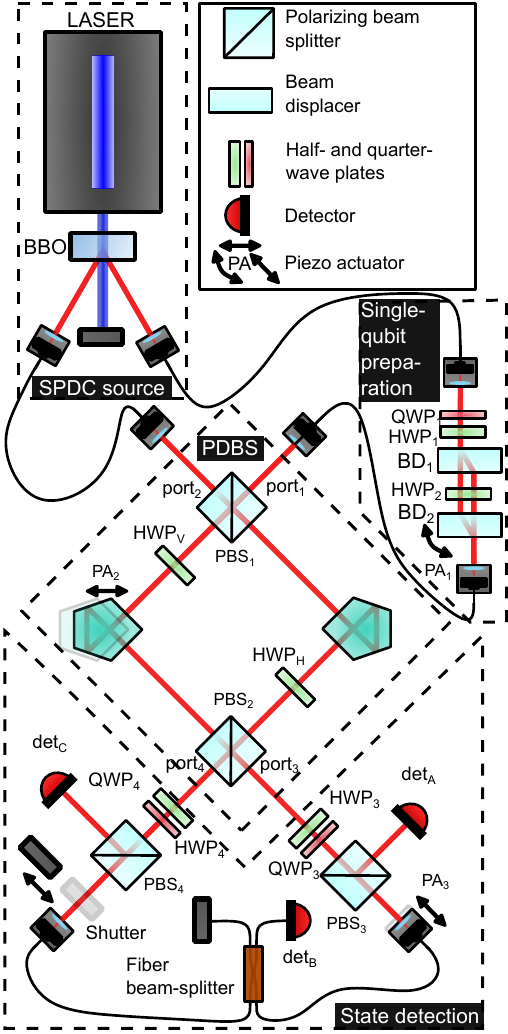}}\hspace*{0pt}\\ %
\caption{Scheme of the experimental setup. SPDC stands for
spontaneous parametric down-conversion, BBO denotes beta barium
borate ($\beta$-BaB\textsubscript{2}O\textsubscript{4}) crystal
and PDBS stands for polarization-dependent beam
splitter.}\label{fig_setup}
 \end{center}
 \end{figure}

We constructed an experimental setup on the platform of linear
optics (see Fig.~\ref{fig_setup}). As mentioned above, we employ
polarization encoding, where the horizontal component of the
polarization represents the single-photon state, while the
vertically polarized component represents the vacuum state.

Separable photon-pairs are generated using a type-I spontaneous
parametric down conversion (SPDC). One photon from each pair is
guided through an optical fiber into the \emph{single-qubit
preparation} stage. There, using a set of motorized quarter-wave
(QWP) and half-wave (HWP) plates, the photon is transformed into
any pure single qubit state in the form of Eq.~(\ref{sigmapx})
with $D=1$. To tune the value of $D$, a set of two beam displacers
(BD) with a half-wave plate rotated by 45° in the middle is used.
The second BD can by tilted by a piezo actuator, allowing us to
set a path difference and, thus, a phase shift ($\Delta\phi$)
between the two polarization components. By changing randomly,
with probability $f$, between the identity ($\Delta\phi=0$) and
the phase flip ($\Delta\phi=\pi$) operations, a desired value of
$D$ is achieved. The single qubit is then guided through another
optical fiber into the first port of the polarization-dependent
beam splitter (PDBS). The second photon from each pair is
vertically polarized, as it plays the role of the vacuum, and is
guided directly from the SPDC source into the second port of the
PDBS.

The PDBS allows us for independent tuning of the beam splitting
ratio for the horizontal and vertical polarization modes. The PDBS
is implemented by means of a Mach-Zehnder interferometer composed
of two polarizing beam splitters (PBSs) with a motorized HWP in
each arm. The horizontal (vertical) polarization components of
both photons meet in arm\textsubscript{H} (arm\textsubscript{V})
after implementing a bit-flip operation to one of the inputs
(HWP\textsubscript{2}).

For example, a vertically polarized photon from
port\textsubscript{2} is transmitted through PBS\textsubscript{1},
while a vertically polarized photon from port\textsubscript{1} is
first transformed, via the bit-flip operation, into a horizontally
polarized photon and subsequently reflected by
PBS\textsubscript{1}. At this point the originally vertical
polarization components of the first and second input modes are
directed into the same spatial mode with the horizontal and
vertical polarizations, respectively. Subsequently, these modes
impinge on HWP\textsubscript{V}, which acts on these polarization
modes identically to a tunable BS acting on spatial modes. Mixing
the polarization states in the arms intertwine the signal from
different inputs, therefore the reflection coefficients for each
polarization component are set by the rotation of the waveplates:
$r_\text{H;V} = \sin(2\theta_\text{H;V})$. The rotation of
HWP\textsubscript{3} is offset by 45° to compensate for the
initial bit-flip operation. For all measurements,
$\theta_\text{V}$ is set to 22.5°, accomplishing a balanced
splitting for the vertical polarization component, corresponding
to the symmetric behavior of the vacuum on a BS, while
$\theta_\text{H}$ is tuned.

Because a HWP rotated by 22.5° serves as a balanced beam-splitter
for the polarization components, the originally vertically
polarized photons, which are directed by PBS\textsubscript{1} into
the same spatial mode as orthogonal polarizations, are subjected
to the Hong-Ou-Mandel effect
$\ket{H}\ket{V}\rightarrow(\ket{H}\ket{H}-\ket{V}\ket{V})/\sqrt{2}$.
Thus, both the vacuum place-holders are bunched at the output of
PBS\textsubscript{2}.

Note that the rotation of HWP\textsubscript{H} enables us to
achieve the optimal splitting ratio for the horizontal
polarization; we set the rotation by 22.5° for balanced
splitting. However in this experiment, we also investigated the
effect of an imperfect interaction upon the studied potentials.
Then, the splitting ratio for the horizontal polarization,
representing the single-photon component, is tuned by rotating
HWP\textsubscript{H}.

Due to the specific nature of the applied encoding, which implies
the bunching at a BS of vertically polarized photons corresponding
to the vacuum, we cannot use standard quantum tomography. We
designed a special detection apparatus and procedure to
reconstruct the output-state density matrix. The detection
apparatus consists of a set of wave plates followed by a
polarizing BS in both output ports of the PDBSs. One output port
of each of these splitters is guided directly to a single-photon
detector, while the other is led to a fiber BS, whose one output
is then guided to a third detector. Using coincidence logic,
events consisting of simultaneous detections by two detectors are
registered.

The reconstructed density matrix takes the form of
\begin{equation} \label{blkMatrix}
\rho_\text{out} =
\begin{pmatrix}
\rho_{11} & \rho_{12} & \rho_{13}& 0\\
\rho_{21} & \rho_{22} & \rho_{23}& 0\\
\rho_{31} & \rho_{32} & \rho_{33}& 0\\
0&0&0&0&
\end{pmatrix}
=
\begin{array}{cccc}
\begin{pmatrix}
M_A & M_C & M_D &0 \\
M_C^*  & \multicolumn{2}{c}{\multirow{2}{*}{\huge $M_B$\normalsize}}&0 \\
M_D^* &  & &0\\
0&0&0&0
\end{pmatrix},
\end{array}
\end{equation}
where \textit{\textbf{M$_B$}} is a $2\times 2$ matrix, and we
assume only the absolute value of the off-diagonal elements,
because any state in Eq.~(\ref{rhoqrpx}) can be transformed into a
fully positive matrix,  using only local rotations and, as a
result, the entanglement measures are independent of the phases in
$\rho_{out}$.

We split the detection procedure into several steps corresponding
to the blocks in the matrix in Eq.~(\ref{blkMatrix}). The term
$M_A$ corresponds to the case of two vertically polarized photons
on the input. Due to the rotation of HWP\textsubscript{V} and the
Hong-Ou-Mandel effect~\cite{Hong1987}, such photons bunch at the
output of the PDBS, meaning they are both in one output port or
the other. Because they propagate symmetrically in both arms and
that the vertical polarization serves only as a placeholder for
the vacuum, we register the signal only in one arm and correct for
the undetected signal accordingly. The term is measured using the
detectors det\textsubscript{A} and det\textsubscript{B}, while the
shutter in port\textsubscript{4} is closed and wave plates in
port\textsubscript{3} are set, so that the bunched photons split
in half of the cases. Overall, the measured signal is multiplied
by the factor of 4 (a factor of 2 by focusing only on one output
arm and another factor of 2 due to the splitting probability).

The term $M_B$ corresponds to the case of one horizontally
polarized photon in either port\textsubscript{3} ($\rho_{22}$) or
port\textsubscript{4} ($\rho_{33}$) and the coherence factor
between these cases ($\rho_{23}$). The vertically polarized
component propagates symmetrically in both output arms. In this
case, we register the vertically polarized component in the output
port where the horizontal component is not detected. To correct
for neglecting the vertical component in the other arm, we
multiply the signal by the factor of 2. The submatrix $M_B$ itself
is measured by standard two-qubit state
tomography~\cite{Jezek2003, Halenkova2012} using
det\textsubscript{A} and det\textsubscript{C} and a maximum
likelihood estimation is applied. This leads to a matrix of the
form
\begin{equation}
\rho_B =
\begin{array}{cccc}
\begin{pmatrix}
0&0&0&0\\
0& \multicolumn{2}{c}{\multirow{2}{*}{\huge $\tilde{M}_B$\normalsize}} &0\\
0&& &0\\
0&0&0&0
\end{pmatrix},
\end{array}
\end{equation}
where $\tilde{M}_B$ is the normalized $M_B$. A proper
renormalization takes into account the measurement of $M_A$.

The terms $M_D$ and $M_C$ in Eq.~(\ref{blkMatrix}) are calculated
directly from the visibility $v$ of the interference pattern,
which is related to the coherence term $v(\rho_{11}+\rho_{33})/2 =
|M_D|$, while the position of the interference pattern corresponds
to the phase of $M_D$. Thus, to acquire the $M_D$ term, we measure
directly the coherence between $\rho_{11}$ and $\rho_{33}$ via an
interference pattern on the fiber BS using the detectors
det\textsubscript{A} and det\textsubscript{B}. In
port\textsubscript{3}, we project onto the term $\rho_{11}$ using
the same setting as when measuring $M_A$. In
port\textsubscript{4}, the shutter is left open, and the
wave-plates are set such that the term $\rho_{33}$ is projected
onto the same set of detectors, leading to the interference of the
two projected terms. A piezo-driven phase shifter
(PA\textsubscript{3}) is used to sweep through the whole
interference pattern, whose maxima and minima can be used to
calculate the absolute value of the coherence term. Note that due
to the instability of the interferometer formed by the
PBS\textsubscript{2} and a fiber BS, it is impossible to precisely
determine the phase of $M_D$. Fortunately, this is unnecessary for
our purposes. On the other hand, a precise determination of the
maxima and minima of an interference pattern is guaranteed by
sweeping multiple times through the pattern. Using the same
measurement method with settings symmetrically swapped for
port\textsubscript{3} and port\textsubscript{4}, we acquire the
value of $M_C$. For more details about the measurement procedure,
see~\cite{Kadlec2024}, where we used this setup for a conceptually
different experiment.

The applied tomography method includes several modifications, when
compared to typical two-qubit tomography schemes (see,
e.g.,~\cite{James2001, Altepeter2005, Miranowicz2014,
Bartkiewicz2016} and references therein). The key difference lies
in the fact that in this experiment, the cases when both photons
leave the PBS\textsubscript{2} at the same output port must be
detected in order to evaluate all the components of a given
density matrix. This is not the case in typical two-qubit
tomography, when measurement is performed solely on the
coincidences across the output ports. This difference constitutes
the reason for the step-by-step reconstruction of the density
matrix from the blocks described in Eq.~(\ref{blkMatrix}).
Specifically, the off-diagonal terms have to be evaluated by
adding a fiber coupler superimposing the output ports. We benefit
from the fact that the phase value of these components does not
have any physical impact on the evaluated entanglement measures.
As a result the interferometer formed by PBS\textsubscript{2} and
this fiber coupler is not required to be phase stable and only the
interference visibility needs to be measured. A maximum likelihood
procedure is used in the final stage of the estimation to ensure
the physicality of the reconstructed density matrices. A similar
procedure requiring an additional fiber coupler and a step-by-step
reconstruction of the density measurement was successfully tested
before, e.g., to verify the preparation of the two-photon
Knill-Laflamme-Milburn states~\cite{KLM2001, Lemr2010}.

At the end of this section, let us also explain the difference
between our current experiment and that reported in
Ref.~\cite{Kadlec2024}: (1)~We are effectively using the old
experimental setup, which however now encompasses an unbalanced
splitting of photons and the possibility of incoherent
interactions. Specifically, the difference is that the wave plate
HWP\textsubscript{H} within the PDBS device is varied to introduce
the incoherent interaction, whereas it remained at a fixed setting
in Ref.~\cite{Kadlec2024}. Thus, tunability was possible with the
previous setup; however, that was not explored since it was not
the goal of our previous experiment. (2)~We perform measurements
of new classes of states. Specifically, the states that have
undergone imperfect interaction, i.e., with an unbalanced
beam-splitting ratio and controllable incoherent interactions.
(3)~We perform a novel postprocessing of the measured states.
Specifically, we apply a probabilistic mixing of the registered
counts for HWP\textsubscript{H} rotated to $\pm 22.5°$ in order
to achieve the incoherent beam splitting.

\section{Experimental results}

\begin{table}
\centering
\begin{ruledtabular}
\begin{tabularx}{\textwidth}{cccccc}
& Class of states & p      & $x$           & $r$   & $w$\\
\hline
(i)&Pure input      & [0,1] & $x_\text{max}$ & [0,1] & 1  \\
(ii)&Dephased input     & [0,1] & 0                 & $1/\sqrt{2}$ & 1 \\
(iii)&Incoherent interaction    & 1 & 0=$x_\text{max}$ &
$1/\sqrt{2}$ & [0,1]
\end{tabularx}
\caption{\label{Table}Classes of experimentally measured states,
where $p$ is the single-photon probability, $x$ is the coherence
parameter, $x_\text{max}=\sqrt{p(1-p)}$ is the maximum value of
the coherence parameter, $r$ is the reflectivity amplitude, and
$w$ characterizes the output state coherence.}
\end{ruledtabular}
\end{table}

\texttt{\begin{figure}
\begin{center}
{\includegraphics[scale=0.88]{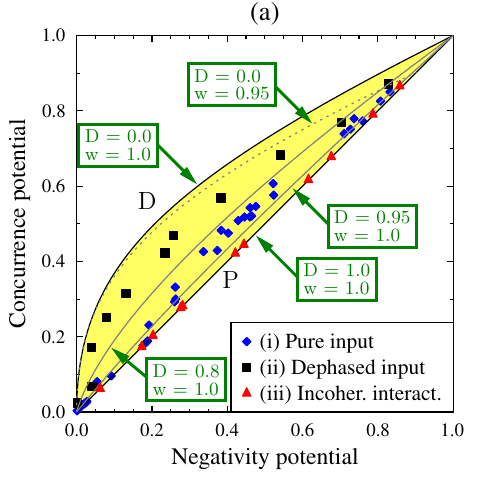}}\hspace*{0pt}\\ %
{\includegraphics[scale=0.88]{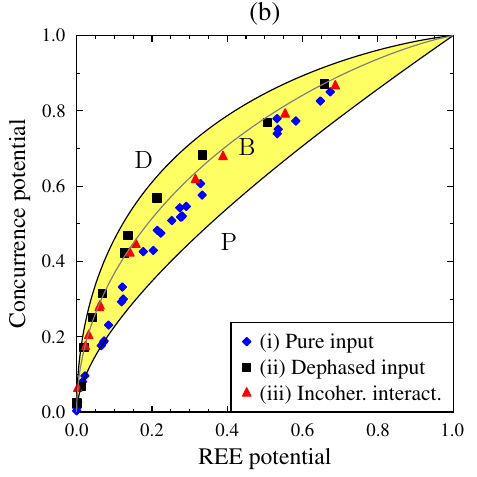}}\hspace*{0pt}\\
{\includegraphics[scale=0.88]{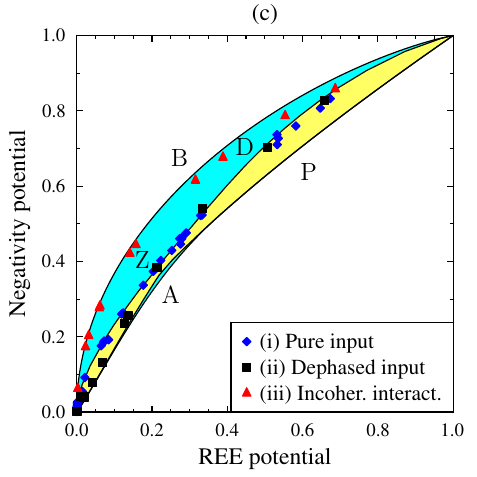}}\hspace*{0pt}
\caption{Relative entanglement potentials defined by various
entanglement measures for the experimental states representing
classes (i)-(iii). The gray middle curve in (c) corresponds to our
theoretical predictions for the states in class
(iii).}\label{fig_exp_bananas}
 \label{neg_conc}
 \end{center}
 \end{figure}}

Here we analyze the following experimentally obtained classes of
states: (i) pure input states and coherent interaction with the
vacuum (for the BS parameters $r\in[0,1]$ and $w=1$), (ii)
dephased input states and coherent interaction with the vacuum,
and (iii) one-photon state subjected to an incoherent interaction
with the vacuum with varying incoherence parameter. For the ranges
of density matrix parameters see Table~\ref{Table}.

To verify that the setup is properly adjusted and the produced
states qualify as class (i), we put a condition on the value of
the off-diagonal terms: $\sqrt{\rho_{12}/\rho_{12}^\text{max}}
\sqrt{\rho_{13}/\rho_{13}^\text{max}} >0.5$, where
$\rho_{1j}^\text{max} = \sqrt{\rho_{11}\rho_{jj}} $ is the maximal
possible value of $\rho_{1j}$ given the values of $\rho_{11}$ and
$\rho_{jj}$ for $j\in \{2,3\}$.

The incoherent interaction needed to prepare the states of class
(iii) was achieved by randomly swapping $\theta_H$ between
$\pm$22.5° and summing the registered coincidence counts.

For all of the output states, we calculated the three entanglement
measures: the negativity, the concurrence, and the REE. All these
quantities were estimated with uncertainties typically below 0.03.
In Fig.~\ref{fig_exp_bananas}, we plotted their mutual relations
for the measured states. In these three plots, the white area
represents mutual relations that are unphysical for any two-qubit
state. In Fig.~\ref{fig_exp_bananas}(a) we show the mutual
relation between the negativity and the concurrence for the
experimental states. This relation allows us to separate dephased
input states from pure input states independently on the
(in)coherence of the interaction. For a better understanding, this
plot depicts theoretical loci for states of various parameters $D$
and $w$.

Figure~\ref{fig_exp_bananas}(a) shows that our experimentally
generated states are located relatively close to the theoretical
lower boundary states (labeled $P$) for any values of NP. Even
though the experimental states dephased at the input [class (ii)]
do not lie exactly on the opposing boundary $D$, they are clearly
separated from the other classes. Moreover,
Fig.~\ref{fig_exp_bananas}(b) shows the relation between the
concurrence and the REE. States of class (iii) do cover the
boundary line $B$ as theoretically predicted. On the other hand,
the experimental states of classes (i) and (ii) do not correspond
exactly to the theoretical boundaries $P$ and $D$, respectively.
Nevertheless, they can be mutually separated by the boundary line
$B$.

Figure~\ref{fig_exp_bananas}(c) demonstrates distinct regions of
physically possible EPs. The yellow inner region represents the
EPs attainable with any input state interacting on a balanced and
lossless BS. The cyan areas are achievable only if the interaction
is incoherent (larger upper area) or the BS is unbalanced (small
lower area). As a result, by analyzing the mutual relation between
the REE and the negativity, we can clearly uncover imperfections
in the interaction. It is seen that states of class (iii) are
located in the cyan region, while all the other states remain in
the yellow region (up to experimental imperfections). Note that
the smaller cyan area on the right-hand side is theoretically
reachable by unbalanced beam splitting. In experiment, this area
proved to be unattainable, because it is very small and that
dephasing of the output (and other experimental imperfections)
pushes the states to the left-hand side of the figure. This
shortcoming, however, does not diminish the main point shown by
this plot. Observing the EPs outside of the yellow area signals
imperfections in the BS interaction and warns that the observed
potentials underestimate the true values of EPs of the tested
states.

Therefore, these relations allow us to distinguish between the
three tested classes of states. The relation between the CP and
the NP distinguishes classes (i) and (iii) from class (ii), while
the relation between the REEP and the NP distinguishes classes (i)
and (ii) from class (iii).

Finally, one can observe in Fig.~\ref{fig_exp_bananas}(c) that our
experimentally generated states are located relatively close to
theoretical boundary states $A$, $B$, $Z$, $D$, and even $P$ for
some ranges of the NP. But we must admit that we failed to
generate high-quality pure states $P$ for $\NP>0.4$. We note that
all the raw and processed experimental data, along with the
processing scripts, are provided in Supplemental Material.
\footnote{See Supplemental Material at https://... for raw and
processed experimental data along with the processing scripts.}

\section{Discussion and Conclusions}

We both theoretically and experimentally analyzed the relations
among the three EPs for various classes of states. These include
the potentials: REEP, CP, and NP. Specifically, we compared CP and
NP, CP and REEP, and NP and REEP for experimentally generated and
tomographically reconstructed states. Our experimental results are
consistent with previous theoretical expectations in the sense of
relative EPs reachable by states belonging to these classes.

We conducted different experiments using essentially the same
optical setup as that used in Ref.~\cite{Kadlec2024}. Therefore,
the main differences from the previous experiment lie in the
results (obtained here for both balanced and unbalanced splitting
of photons) and their interpretation. Specifically, we
experimentally generated single-qubit states close to those with
maximal and minimal relative EPs. Note that Ref.~\cite{Kadlec2024}
focused on showing the feasibility of our experimental setup for
demonstrating the hierarchy of the potentials for quantum
entanglement, EPR steering, and Bell nonlocality.

In our experiments, we employed photon polarization instead of
photon-number encoding, which is typically considered for the
determination of EPs. Specifically, we assumed that the horizontal
and vertical components of polarization states represent the
single-photon and vacuum states, respectively. We emphasize again
that the applied polarization encoding serves as an effective
analog to the encoding of VOPS states, significantly simplifying
their physical realization. Indeed, polarization encoding is much
simpler than photon-number encoding or single-rail encoding, which
are affected by photon loss and dark counts. Nevertheless, the use
of polarization encoding does come with its own experimental
challenges. Most notably, an imperfect polarization adjustment
using polarization encoding may manifest as photon creation in the
framework of the VOPS states. We minimize this adverse effect by
fine tuning of the setup and by adding polarizers to the input,
thus purifying the polarization states obtained from the
two-photon source. It should be pointed out that a similar problem
arises in the genuine VOPS encoding considering that single-photon
detection is burdened by dark detection events. Photon loss that
cannot be easily prevented poses a significant challenge to the
implementation of VOPS states. This issue does not affect the
results obtained in the polarization encoding because of the
postselection on two-photon detection events. Thus, we can
overcome the majority of technological losses which are
polarization independent, including fiber coupling,
back-reflection, absorption on imperfect components, and detector
efficiency.

Based on the experimentally reconstructed two-qubit states
generated from single-photon states, we determined the EPs defined
by the Peres-Horodecki negativity, the Wootters concurrence, and
the relative entropy of entanglement. Among our experimental
states we were able to find those, which are very close to the
theoretical upper and lower bounds of the relative EPs, as
predicted in Refs.~\cite{Miranowicz2015b, Miranowicz2015,
Miranowicz2023}.

Moreover, we confirmed experimentally the counterintuitive
theoretical result of Ref.~\cite{Miranowicz2015} that the NP for
given values of the REEP can be increased by dissipation. We note
that our improved experimental setup, in contrast to that of
Ref.~\cite{Kadlec2024} enables performing unbalanced splitting of
photons and the possibility of introducing controllable incoherent
interactions. This is accompanied by measured-states
postprocessing, which was not used in the previously reported
experiment.

We documented the benefits of studying the relations between EPs,
as they allow to detect imperfect interaction between the tested
state and the classical state. Detecting imperfections in the
interaction is critical in preventing misjudging the states' true
nonclassicality. Considering that imperfections are unavoidable in
experimental reality, especially in near-future quantum
technologies, we believe that our findings are relevant for the
practical deployment of EPs as a method for nonclassicality
quantification.

Further analysis could be performed to allow us to establish
markers of various interaction flaws (unbalance beam splitting,
decoherence, and amplitude damping). One could, for instance,
immediately deduce the incoherence of the interaction by comparing
the absolute value of the term $\rho_{12}$ to its maximum value
$\rho_{12}^{\text{max}}$. Our analysis, however, enables us to
estimate the impact of this incoherence on the observed EPs.

\section*{Acknowledgements}

J.K. gratefully acknowledges the support from the Project
IGA\_PrF\_2024\_004 of Palack\'y University. A.Č. and K.L.
acknowledge support from the Project OP JAC
CZ.02.01.01/00/22\_008/0004596 of the Ministry of Education,
Youth, and Sports of the Czech Republic. K.B. and A.M. were
supported by the Polish National Science Centre (NCN) under
Maestro Grant No. DEC-2019/34/A/ST2/00081. We thank Cesnet for
data management services.

\appendix

\section{Maximally nonclassical single-qubit states}

Here we recall, after Refs.~\cite{Miranowicz2015,
Miranowicz2015b}, definitions of maximally nonclassical
single-qubit states and their corresponding two-qubit states,
which are located at the upper or lower theoretical bounds of
various relative EPs, as shown in Figs.~2 and~4. Note that the
Secs. $a$ to $d$ define the boundary state for the yellow area in
Figs.~2 and~4, while the remaining sections define the boundary
states for the extended cyan area.

\subsubsection{Pure states $\SIGMA{P}$}

First we consider single-qubit pure states,
\begin{equation}
  \ket{\psi_p} = \sqrt{1-p}\ket{0}+e^{i\phi}\sqrt{p}\ket{1},
 \label{psi_p}
\end{equation}
which are nonclassical for any  $p\in(0,1]$. Note that
$\SIGMA{P}\equiv\ket{\psi_p}\bra{\psi_p}$ is a special case of
Eq.~(\ref{sigmapx}) for $|x|=\sqrt{p(1-p)}=x_\text{max}$ and
$\phi={\rm Arg}(x)$. An arbitrary pure state $\ket{\psi_p}$ after
being mixed with the vacuum at the balanced and lossless BS is
transformed into the entangled states for $p\neq 0$:
\begin{equation}
|\Psi_{\rm out}(p) \rangle
=\sqrt{1-p}|00\>+\sqrt{\tfrac{p}{2}}(|10\>-|01\>),
\label{PsiRhoOut}
\end{equation}
which reduces to the singlet state for $p=1$. The EPs are simply
given by:
\begin{eqnarray}
\REEP(\SIGMA{P}) &=& h\big(\textstyle{\frac{1}{2}}[1+\sqrt{1-p^2}]\big),\nonumber \\
 \CP(\SIGMA{P}) &=& \NP(\SIGMA{P})=p, \label{REE_P}
\end{eqnarray}
where $h$ is the binary entropy.

As seen in Fig.~2, single-qubit pure states are the maximally
nonclassical states corresponding to the upper bounds for: (1) the
NP and (2) the REEP as functions of CP$\in[0,1]$, and (3) the REEP
as a function of NP$\in[N_2,1]$, where $N_2 \approx 0.527$.

Moreover, on the scale of Fig.~2(c), we cannot see any differences
between the curves for the pure states $\sigma_P$ and the
optimally dephased states $\sigma_Z$, as defined below, if
$\NP<N_0\approx 0.2$. Experimentally these $\sigma_P$ and
$\sigma_Z$ cannot be distinguished for this range of NP. Thus,
effectively, pure states can also be considered maximally
nonclassical in terms of: (4) the largest NP as a function of the
REEP for NP in $\in[0,N_0]$ assuming a balanced and lossless BS.

\subsubsection{Completely dephased states $\SIGMA{D}$}

Completely dephased single-qubit states, which are also referred
to as completely mixed states~\cite{Miranowicz2015,
Miranowicz2015b}, are the mixtures of $\ket{0}$ and $\ket{1}$
corresponding to the special case of Eq.~(\ref{sigmapx}) for the
vanishing coherence parameter $x=0$, i.e.,
\begin{equation}
  \SIGMA{D} = \sigma(p,x=0)= (1-p)\ket{0}\bra{0}+p\ket{1}\bra{1}.
 \label{RhoM}
\end{equation}
We recall that $\SIGMA{D}$ is transformed by the balanced and
lossless BS (with the vacuum in the other port) into the Horodecki
state,
\begin{equation}
  \RHO{H}(p) = \rho_{\rm out}(p,0) =
  p\ket{\Psi^-}\bra{\Psi^-}+(1-p)\ket{00}\bra{00},
 \label{rhoH}
\end{equation}
which is A mixture of the singlet state,
$\ket{\Psi^-}=(\ket{10}-\ket{01})/\sqrt{2}$, and the vacuum. The
entanglement (together with EPR steering and Bell nonlocality) of
the two-photon Horodecki states has been studied intensively (see,
e.g., Refs.~\cite{Horodecki09review, Jirakova2021, Abo2023} and
references therein). The EPs for $\rho_{\rm D}$, thus, correspond
to the known entanglement measures of the Horodecki
states~\cite{Horst2013}:
\begin{eqnarray}
\REEP(\rho_{\rm D}) &=&
(p-2)\log_{2}(1-\tfrac{p}2)+(1-p)\log_{2}(1-p),\nonumber \\
  \NP(\rho_{\rm D}) &=& \sqrt{(1-p)^2+p^2}-(1-p),\label{REE_H}
\end{eqnarray}
and  $\CP(\rho_{\rm D})=p$.

The completely dephased states $\rho_{\rm D}$, as shown in Fig.~2,
are the maximally nonclassical single-qubit states with respect to
the largest values of: (1) the CP as a function of $\NP\in[0,1]$,
(2) the CP vs. $\REEP\in[0,1]$, (3) the REEP vs.
$\NP\in[0,N_{2}]$, where $N_0\approx 0.2$; and (4) assuming a
balanced and lossless BS, the NP vs. REEP $\in[E_3,1],$ where
$E_3=0.397$.

\subsubsection{Optimally dephased states $\SIGMA{Z}$}

Optimally dephased single-qubit states, which maximize the NP for
a given value of the REEP assuming a perfectly balanced and
lossless BS in an EP setup, are defined as~\cite{Miranowicz2015}:
\begin{eqnarray}
  \SIGMA{Z}(\bar{N}) &=&\sigma[p_{\rm opt},x_{\rm opt}=f(p_{\rm
  opt},\bar{N})], \label{rhoX}
\end{eqnarray}
where for brevity we denote $\bar{N}\equiv \NP$
and\begin{equation}
  f(p,\bar{N})= \tfrac12 \sqrt{(1+p/\bar{N}) [2\bar{N}(\bar{N}+1)-(\bar{N}+p)^2]}.
 \label{f}
\end{equation}
The optimal probability $p_{\rm opt}$ is found numerically by
minimizing
\begin{equation}
  \REEP\{\sigma[p_{\rm opt},f(p_{\rm opt},\bar{N})]\}=\min_p
\REEP\{\sigma[p,f(p,\bar{N})]\}.
  \label{min_REEP}
\end{equation}
Here, the minimalization is performed for $p\in[\bar N,\sqrt{2\bar
N(\bar N+1)}-\bar N]$ for a given $\bar N$. It is seen in
Fig.~2(c) that $\SIGMA{Z}(\bar{N})$ is practically
indistinguishable from $\SIGMA{D}$ for $\bar{N}>N_3\approx 0.6$
or, equivalently, for $\REEP>E_3\approx 0.397$. Moreover,
$\SIGMA{Z}(\bar{N})$ goes into pure states $\SIGMA{P}$ for small
$\bar{N}$ (say, $\bar{N}\lesssim 0.2$). Thus, these partially
dephased states become completely dephased for large EPs and
completely purified for small EPs. However, $\SIGMA{Z}(\bar{N})$
is clearly different from both $\SIGMA{P}$ and $\SIGMA{D}$ for
$\bar{N}$ close to $N_1\approx 0.377$.

\subsubsection{Optimally dephased states $\SIGMA{Y}$}

Note that there exist optimally dephased single-qubit states (say,
$\SIGMA{Y}$), when comparing the NP and the REEP in Fig. 2(c), for
which $\NP(\SIGMA{Y})$ has slightly lower values than both
$\NP(\SIGMA{D})$ and $\NP(\SIGMA{P})$, especially near the
crossing at $N_1\approx 0.377$ of the curves for $\NP(\SIGMA{D})$
and $\NP(\SIGMA{P})$. Nevertheless, the states $\NP(\SIGMA{Y})$
cannot be distinguished from $\NP(\SIGMA{D})$ (if $\NP\le N_1$)
and $\NP(\SIGMA{D})$ (if $\NP\ge N_1$) on the scale of Fig. 2(c),
therefore, due to experimental uncertainty, they also cannot be
distinguished experimentally. Thus, $\SIGMA{Y}$ are not discussed
in detail in this paper, and consequently
$\min[\NP(\SIGMA{D}),\NP(\SIGMA{P})]$ are considered to be an
effective lower bound for the NP vs. REEP assuming a balanced and
lossless BS in our setup.

\subsubsection{Boundary states $\RHO{A}$ by unbalanced beam splitting}

The above four classes of two-qubit states can be generated from
single-qubit states assuming a perfectly balanced and lossless BS
($\theta=\pi/2$) in an ideal EP-detection scheme. Here we assume
that the BS can be tuned to change its reflectivity,
$R=r^2=\sin^2(\theta/2)$, and transmissivity,
$T=t^2=\cos^2(\theta/2)$. Then, a completely dephased state
$\SIGMA{D} = \sigma(p,x=0)$ is transformed by a tunable BS into
the generalized Horodecki state~\cite{Miranowicz2015}:
\begin{eqnarray}
  \rho^{\theta}_{\rm out}(p,x=0)
  &=&p |\Psi_q\> \< \Psi_q| + (1-p)|00\>\<00| \nonumber \\
  &\equiv& \RHO{GH}(p,q=R),
 \label{GH_TBS}
\end{eqnarray}
where $p,q\in [0,1]$ and
\begin{equation}
  |\Psi_q \rangle =\sqrt{q} |01\rangle - \sqrt{1-q} |10\rangle.
  \label{Psi_q}
\end{equation}
The boundary states $\rho_A$, which are shown in Figs. 2(c) and
4(c), are given as
\begin{equation}
  \RHO{A}(\bar{N})=\RHO{GH}[\bar p_{\rm opt},\bar q_{\rm opt}],\quad
 \label{rhoA}
\end{equation}
where $\bar q_{\rm opt}=f_1(\bar p_{\rm opt},\bar{N})$, and
\begin{equation}
  f_1(p,\bar{N}) = \frac{1}{2p}\left[p\pm \sqrt{p^2-\bar{N}^2-2\bar{N}(1-p)}\right],
\label{f_1}
\end{equation}
for the optimized value of the mixing parameter $\bar p_{\rm
opt}(\bar{N})$, which can be found numerically such that
\begin{eqnarray}
 & {\rm REE}\{\RHO{GH}[\bar p_{\rm opt},\bar q_{\rm opt})]\}\hspace{3cm} \nonumber\\
 & \quad \quad\quad =\max_p {\rm REE}\{\RHO{GH}[p,f_1(p,\bar{N})]\}.
 \label{p_opt}
\end{eqnarray}
Here, the maximization is performed for such $p$ given $\bar N$
that $f_1(p,\bar N)\in [0,1]$.

\subsubsection{Boundary states $\RHO{A}$ by amplitude damping}

Here we describe another method for generating the boundary states
$\RHO{A}$, where instead of unbalanced BS, as described above, we
use again a balanced BS, but allow for amplitude damping.

A single-qubit pure state $\ket{\psi_p}$ is transformed by a
balanced BS into $\ket{\Psi_{\rm out}(p)}$, given by
Eq.~(\ref{PsiRhoOut}). By applying local unitary transformations,
$\ket{\Psi_{\rm out}(p)}$ can be converted into $\ket{\Psi_q}$
\cite{Nielsen1999}, given by Eq.~(\ref{Psi_q}), where
$q=(1-\sqrt{1-p^2})/2$, without changing its entanglement, as
$C(\ket{\Psi_{\rm out}}) =C(\ket{\Psi_q})=2\sqrt{q (1 - q)}=p$.

Let us assume now that each qubit ($i=1,2$) in $|\Psi_q \rangle$
undergoes amplitude damping, as described by the standard Kraus
operators~\cite{Nielsen2010}:
\begin{equation}
E_{0}(\gamma_{i})=|0\rangle\langle0|+\sqrt{1- \gamma_{i}}
|1\rangle\langle1|,\quad
E_{1}(\gamma_{i})=\sqrt{\gamma_{i}}|0\rangle\langle1|,\label{Kraus_adc}
\end{equation}
with $\gamma_{i}$ being amplitude-damping coefficients. Then a
pure state $|\Psi_{q}\rangle$ for any $q\in [0,1]$ is transformed
into the generalized Horodecki state~\cite{Horst2013,
Bartkiewicz2013}:
\begin{eqnarray}
\RHO{ADC}(q,\gamma_{1},\gamma_{2})&=& \RHO{GH}(p',q')  \label{rho_adc}\\
&=& p' |\Psi_{q'}\> \< \Psi_{q'}| + (1-p')|00\>\<00|,\quad
\nonumber
\end{eqnarray}
where $p'=1-(1-q)(1-\gamma_{1})-q(1-\gamma_{2})$ and $q'=q
(1-\gamma_{2})/(1-p')$. Thus, to generate a boundary state
$\rho_A$, we optimize $p'$ according to
Eqs.~(\ref{rhoA})--(\ref{p_opt}), where $p$ and $q$ should here be
replaced by $p'$ and $q'$, respectively.

We conclude that $\rho_A$ can be considered maximally nonclassical
two-qubit states in terms of the largest REE for a given value of
the negativity assuming $N\in(0,N_{2})$ or effectively for
$N\in(N_0,N_{2})$ for $N_0\approx 0.2$. Because $\rho_A$ can be
generated from $\ket{\psi_p}$  by amplitude damping and from
$\SIGMA{D}$ by an unbalanced BS, we can consider these two classes
of states to be related to the maximally nonclassical single-qubit
states in terms of the largest REEP as a function of the NP for
$\bar{N}\in(0,N_2)$ or, clearly, for $\bar{N}\in(N_0,N_2)$.

\subsubsection{Boundary states $\RHO{B}$ by phase damping}

Here we show after Refs.~\cite{Horst2013, Miranowicz2015} how to
generate the boundary states $\rho_{B}$ from $\ket{\psi_p}$, first
by transforming it to $|\Psi_{\rm out}(p)\rangle$ and then to
$\ket{\Psi_q}$, and finally applying phase damping. The method is
analogous to generating $\rho_{A}$ by amplitude damping, which is
here replaced by phase damping described the Kraus
operators~\cite{Nielsen2010}:
\begin{equation}
E_{0}(\kappa_{i})=|0\rangle\langle0|+\sqrt{1-\kappa_{i}}|1\rangle\langle1|,\quad
E_{1}(\kappa_{i})=\sqrt{\kappa_{i}}|1\rangle\langle1|,\label{Kraus_pdc}
\end{equation}
with $\kappa_i$ being phase-damping coefficients with $i=1,2$.
Thus, a pure state $|\Psi_q\rangle$ is changed into the mixed
state~\cite{Horst2013}:
\begin{eqnarray}
\RHO{PDC}(q,\kappa_{1},\kappa_{2})=(\textstyle{\frac{1}{2}}-y)|\beta_{1}\rangle\langle\beta_{1}|
+(\textstyle{\frac{1}{2}}+y)|\beta_{2}\rangle\langle\beta_{2}|\nonumber \\
 +(q-\textstyle{\frac{1}{2}})(|\beta_{1}\rangle\langle\beta_{2}|+|\beta_{2}\rangle\langle\beta_{1}|),
\hspace{5mm}
\end{eqnarray}
given in the Bell-state basis, where
$\ket{\beta_{1,2}}=\ket{\Psi_{\mp}}
=(\ket{10}\mp\ket{01})/\sqrt{2}$, and
$y=\sqrt{q(1-q)(1-\kappa_{1})(1-\kappa_{2})}$. If we assume the
input state $\ket{\psi_{p=1}}=\ket{1}$, which corresponds to
setting $q=1/2$, then $\RHO{PDC}$ becomes the following
Bell-diagonal state
\begin{eqnarray}
\RHO{B}(\kappa_{1},\kappa_{2})&=&\RHO{PDC}(\tfrac12,\kappa_{1},\kappa_{2})
\nonumber\\ &=&\lambda_-|\beta_{1}\rangle\langle\beta_{1}|
+\lambda_{+}|\beta_{2}\rangle\langle\beta_{2}|, \hspace{5mm}
\label{rhoB1}
\end{eqnarray}
where $\lambda_{\pm}=[1\pm\sqrt{(1-\kappa_{1})(1-\kappa_{2})}]/2$.
The states $\RHO{B}(\kappa_{1},\kappa_{2})$ for
$\kappa_{i}\in[0,1]$ are the examples of the boundary states
labeled `$B$' in Figs.~2 and 4.

Indeed, the states $\rho_B$ can be interpreted as the maximally
nonclassical two-qubit states in terms of the largest negativity
for a given value of the REE~\cite{Miranowicz2004b, Horst2013}.
Since $\rho_B$ can be generated from $\ket{1}$, we produced
experimentally the maximally nonclassical single-qubit states in
terms of the largest (generalized) NP as a function of the
(generalized) REEP in almost the entire range [0,1], as shown in
Fig.~4(c).

\end{document}